\documentclass[12pt,preprint]{aastex}

\newcommand{\G}{G75.30+1.32}
\newcommand{\kms}{km s$^{-1}$}
\newcommand{\HI}{\hbox{H~{\sc i}}}
\newcommand{\HII}{\hbox{H~{\sc ii}}}

\shorttitle{Trigonometric Parallaxes: G75.30+1.32}

\begin{document}

\title{TRIGONOMETRIC PARALLAXES OF MASSIVE STAR-FORMING REGIONS. IX. THE OUTER ARM IN THE FIRST QUADRANT}

\author{A. Sanna\altaffilmark{1}, M. J. Reid\altaffilmark{2},
T. M. Dame\altaffilmark{2}, K. M. Menten\altaffilmark{1},
A. Brunthaler\altaffilmark{1}, L. Moscadelli\altaffilmark{3},
X. W. Zheng\altaffilmark{4} and Y. Xu\altaffilmark{5}}

\email{asanna@mpifr-bonn.mpg.de}
\altaffiltext{1}{Max-Planck-Institut f\"{u}r Radioastronomie, Auf dem H\"{u}gel 69, 53121 Bonn, Germany}
\altaffiltext{2}{Harvard-Smithsonian Center for Astrophysics, 60 Garden Street, Cambridge, MA 02138, USA}
\altaffiltext{3}{INAF, Osservatorio Astrofisico di Arcetri, Largo E. Fermi 5, 50125 Firenze, Italy}
\altaffiltext{4}{Department of Astronomy, Nanjing University, Nanjing 210093, China}
\altaffiltext{5}{Purple Mountain Observatory, Chinese Academy of Sciences, Nanjing 210093, China}

\begin{abstract}

We report a trigonometric parallax measurement with the Very Long Baseline Array for the water maser
in the distant high-mass star-forming region \G. This source has a heliocentric distance of
$9.25^{+0.45}_{-0.40}$~kpc, which places it in the Outer arm in the first Galactic quadrant.
It lies 200~pc above the Galactic plane and is associated with a substantial \HI\ enhancement at the
border of a large molecular cloud. At a Galactocentric radius of 10.7~kpc, \G\ is in a region
of the Galaxy where the disk is significantly warped toward the North Galactic Pole.
While the star-forming region has an instantaneous Galactic orbit that is nearly circular, it displays a
significant motion of 18~\kms\ toward the Galactic plane. The present results, when combined with two
previous maser studies in the Outer arm, yield a pitch angle of about $12\degr$ for a large section of
the arm extending from the first quadrant to the third.

\end{abstract}


\section{Introduction}

The section of the Galaxy beyond the solar circle in the first quadrant is arguably the most
difficult in which to study Galactic structure. Heliocentric distances are large (e.g.,
\citealt{Mead1988,Digel1990}), Population~{\sc i} objects scarce, the rotation curve poorly constrained,
and the disk warped and flared. Still, two well-defined spiral arms are easily recognized in large-scale
21~cm and CO surveys of the region (e.g., \citealt{Dame2001}; see also our Figure~\ref{fig4} below). The
nearer one, the Perseus arm, straddles the solar circle in the first quadrant, apparently crossing it near
a longitude of $\sim 50\degr$; the other is 2--3 kpc farther from the Galactic center and is generally called
the Outer arm (although other names such as "Cygnus-Outer", "Norma-Cygnus", and "Perseus+I" have  been applied;
\citealt{Vallee2008}). We note in passing that an even more distant arm was recently identified in both 21~cm
and CO emissions in the first quadrant \citep{Dame2011}; at about 15~kpc from the Galactic center this may be
the distant end of the Scutum-Centaurus arm.

In general, the Outer arm contains much less molecular gas than arms at smaller Galactic radii (e.g., Perseus).
Even so, some star-forming regions are detected in the arm at distances of $\sim5$~kpc toward the anticenter direction, and
this section of the arm has been located by trigonometric parallaxes \citep{Honma2007,Hachisuka2009}, through
kinematic distance estimates (e.g., \citealt{Russeil2007}), and via photometric observations of open clusters
(e.g., \citealt{Pandey2006}). An accurate knowledge of the gas distribution and large-scale motions in the outer
regions of the Galaxy may provide strong constraints on the rotation curve of the Milky Way. In this respect,
trigonometric parallax measurements of high-mass star-forming regions (HMSFRs) yield both an accurate distance
to the source (within 10\% accuracy) and its full-space proper motion around the Galaxy (within a few \kms). So far, these
observations have provided evidence for a nearly flat rotation curve up to about 13~kpc from the Galactic center in
both the third \citep{Honma2007} and second Galactic quadrants \citep{Hachisuka2009,Reid2009b}.

In this paper, we further constrain the outer-Galaxy rotation curve and the structure of the Outer arm in the first
quadrant using masers associated with the HMSFR \objectname{G75.30+1.32} (IRAS~201444+3726). The longitude and large
negative local standard of rest (LSR) velocity of this region, $-57$~\kms, place it within the locus of the Outer
arm as traced by the gas. We conducted multi-epoch, astrometric, Very Long Baseline Array (VLBA)
observations of its 22~GHz water masers ($>50$~Jy; \citealt{Brand1994}) in order to measure the trigonometric
parallax and 3-dimensional velocity vectors of this star-forming region.

\section{Observations and Data Analysis}

We conducted VLBA\footnote{The VLBA is operated by the National Radio Astronomy Observatory (NRAO).
The NRAO is a facility of the National Science Foundation operated under cooperative agreement by
Associated Universities, Inc.} observations to study the $6_{16}-5_{23}$ H$_2$O
maser emission (rest frequency 22.235079~GHz) toward the HMSFR \objectname{G75.30+1.32}.
In order to measure the trigonometric (annual) parallax and Galactic proper
motion of this source, we used phase-referencing observations by fast switching between the
maser target and two extragalactic continuum sources, J2015+3710 and J2018+3812.
The former one was found in the VCS2 survey and has an absolute position
known to better than $\pm1$~mas \citep{Fomalont2003}, whereas the latter belongs to the VERA
22~GHz Fringe Search Survey (\citealt{Petrov2007}; see Figure~\ref{calib}).
Two strong fringe-finders (3C345 and 3C454.3) were observed for bandpass, single-band delay,
and instrumental phase-offset calibration. We also employed four blocks of
geodetic-like observations, in order to remove total atmospheric delays for each antenna \citep{Reid2009a}.
Detailed source information is summarized in Table~\ref{tab1}.

\begin{figure*}
\centering
\includegraphics[angle= 0, scale= 0.4]{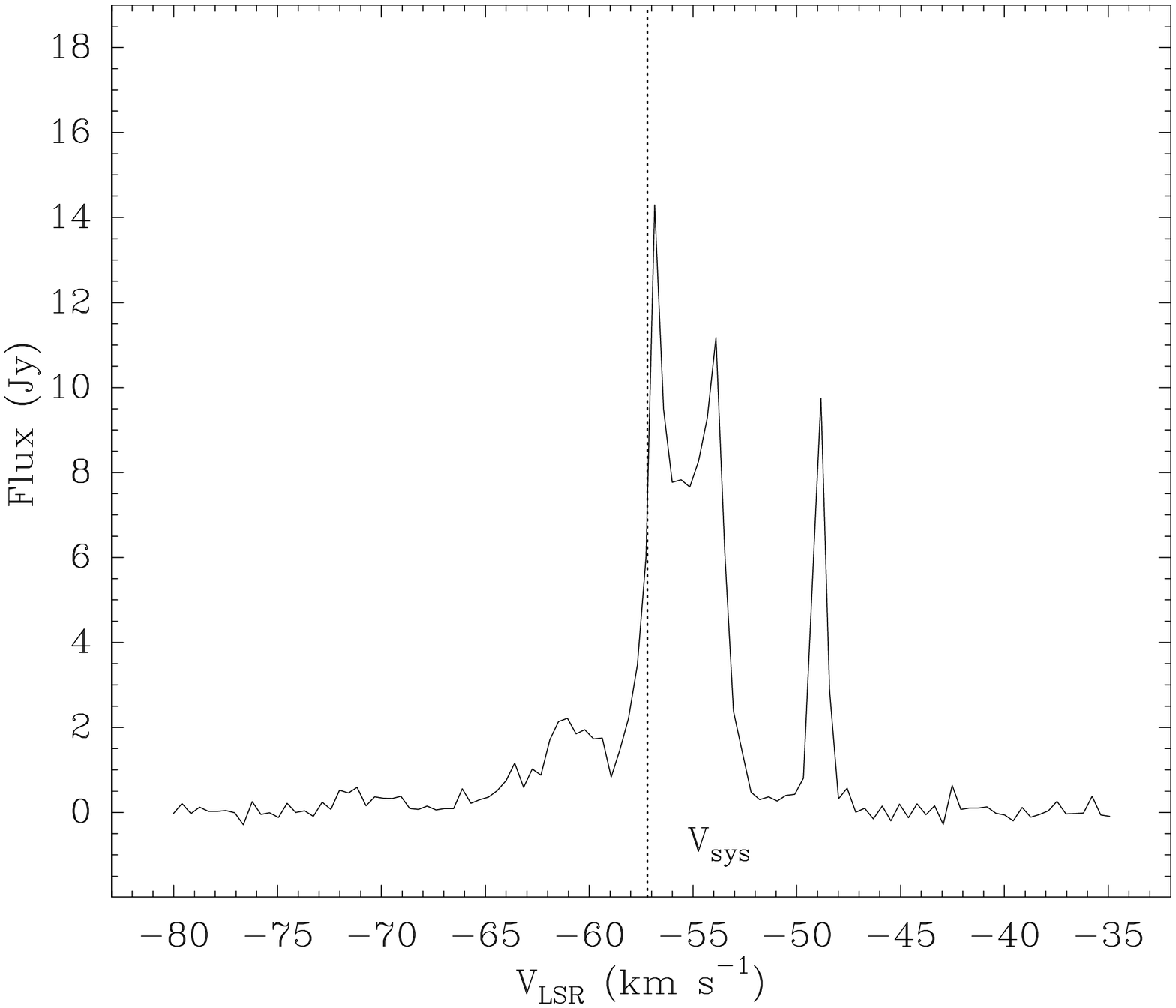}
\caption{Total-power (Stokes~I) spectrum of the 22~GHz H$_2$O masers toward G75.30+1.32 from the second epoch on 2009 May 6. This profile was
produced by averaging the total-power spectra of all VLBA antennas, after weighting each spectrum with the antenna system temperature. The dotted
line crossing the spectrum (at $\rm -57.0~km~s^{-1}$) represents the systemic velocity (V$_{sys}$) of the molecular
cloud hosting the star-forming region \citep{Bronfman1996}. \label{spectrum}}
\end{figure*}

\begin{figure*}
\centering
\includegraphics[angle= -90, scale= 0.7]{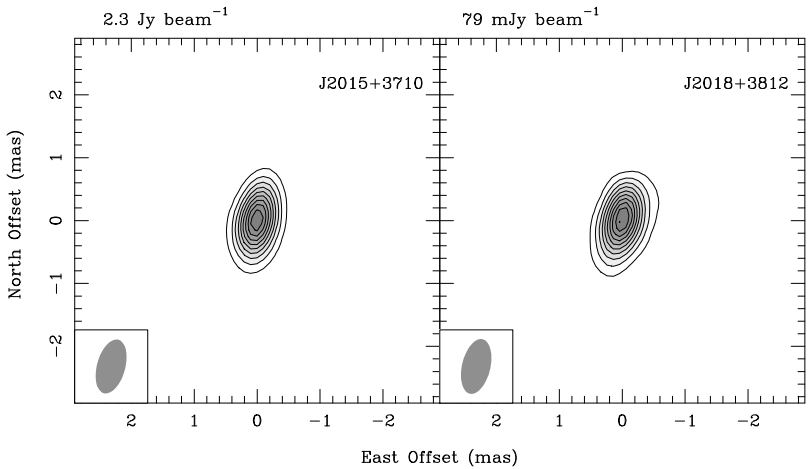}
\caption{Images of the background continuum sources in K--band
near G75.30+1.32 used for parallax purposes. Source names are in the upper right corner and restoring beams are in the lower left
corner of each panel. Contour levels are at multiples of 10\% of each peak brightness reported on top of each panel. All images are
from the second (middle) epoch observations on 2009 May 6 (see Table~\ref{tab1}). \label{calib}}
\end{figure*}

At first, we carried out a preparatory survey with the VLA in BnA-configuration under program BM272 on
2007 October 5, in order to determine the maser position with sub-arcsecond accuracy.
The VLBA observations were scheduled under program BM272H at four epochs: 2008 November 10, 2009 May 6,
13, and November 13.  These dates were optimized to sample the yearly peaks of the right-ascension parallax
sinusoid and at the same time remove all correlation among the parallax and proper motion parameters
(e.g., \citealt{Sato2010}).  We employed four adjacent intermediate frequency (IF) bands, each 8 MHz wide,
in dual circular polarization; each band was correlated to produce 256 spectral channels.
The bandwidth was wide enough to detect even weak continuum sources (a few \ mJy beam$^{-1}$) and the channel
width of 0.42~\kms\ was narrow enough to resolve well the masers line's spectral components.
The third IF band was centered on an LSR velocity (V$_{\rm LSR}$) of --56.0~\kms, covering the range of
previously detected water maser emission \citep{Brand1994}, whereas the remaining three IFs were
spaced by 108~\kms\ each one from the other.  The data were processed with
the VLBA correlation facility in Socorro (New Mexico) using an averaging time of about 0.9~s,
which limited the instantaneous field of view of the interferometer to about $2''$
(i.e. without significant amplitude losses).  Data were reduced with the NRAO Astronomical Image Processing
System (AIPS) following the procedure described in \citet{Reid2009a}.  A total-power
spectrum of the 22.2~GHz masers toward \G\ is shown in Figure~\ref{spectrum}.

\section{Results}

\subsection{Maser emission}

We mapped a range of LSR velocities within $\pm 30$~\kms\ about the mean velocity of the emission ($\approx 56$~\kms)
and a field of view of $\pm 1''$ about the peak position of the reference feature No.~2 (see Table~\ref{tab3}).
Since \G\ has not been well studied in the past, we imaged a large field-of-view and checked for differences
between the total-power spectrum (Figure~\ref{spectrum}) and the flux density from the detected maser features (Table~\ref{tab3}).
With the interferometer, we recovered more than $90\%$ of the emission detected with single-dish measurements.
In addition to the emission reported in Table~\ref{tab3}, we found a strong, maser line at about $-49$~\kms\
in the spectra of the second and third epochs.  This emission came from about $6''$ northward of the reference
feature and probably represents another forming star in the region.
However, analysis of this source is beyond the scope of this paper.

We identified 38 distinct water maser features, distributed within an area of about $ 0\farcs6 \times 0\farcs9 $.
We use the term \emph{feature} to refer to multiple ``spots'' spatially overlapping in contiguous
velocity channels and identify features with individual masing cloudlets (e.g., \citealt{Sanna2010a}).
We used the AIPS task JMFIT to fit an elliptical Gaussian brightness distribution to the brightness
structure of each maser emission center at each epoch.
The measured properties of individual features are presented in Table~\ref{tab3}.
Their intensities range from about 10 to 0.1~Jy~beam$^{-1}$ and span LSR velocities from $-48$~\kms\ for the
most redshifted feature (No.~17) to $-66$~\kms\ for the most blueshifted one (No.~18).
One-third of the maser features persisted over the 1~year time-baseline of our observations.
For features lasting at least 6 months, we modeled their relative position shifts with time by a linear
fit to get the internal proper motions of the maser distribution.
At a distance of 9.25~kpc, the magnitude of the relative motions with respect to
feature No.~2 ranges from 4~\kms\ for feature No.~6 to 84~\kms\ for feature No.~34,
with an average accuracy of about 2~\kms.

\subsection{Distance and Galactic proper motion}\label{distance_results}

\begin{figure*}
\centering
\includegraphics[angle=-90,scale=.6]{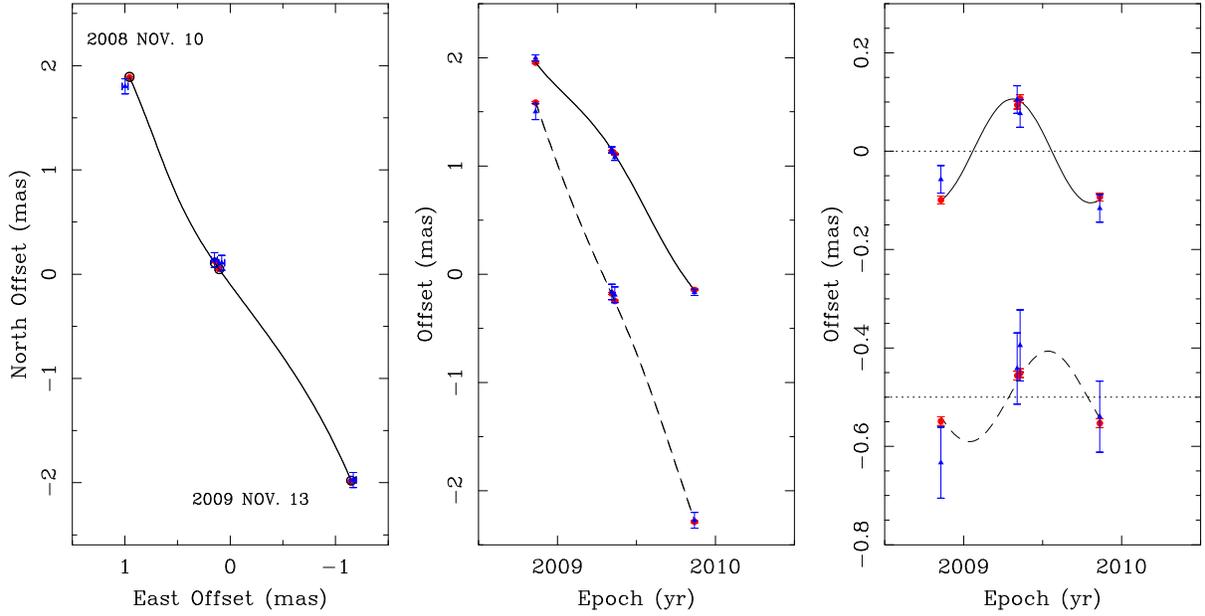}
\caption{Results of the ``combined'' parallax fit for G75.30+1.32. Red circles and blue triangles show maser positions measured
relative to J2015+3710 and J2018+3812, respectively. \textit{Left Panel:} Sky projected motion of the maser with respect
to J2015+3710 and J2018+3812 with first and fourth epochs labeled. The empty circles and the line show the best-fit position
offsets and the trajectory, respectively. \textit{Middle Panel:} The position offsets of the maser along the east and north
directions versus time. The best-fit model in east and north directions are shown as continuous and dashed lines,
respectively. \textit{Right Panel:} Same as the middle panel but with fitted proper motions subtracted (i.e. parallax curve).
The north offset data have been shifted for clarity. \label{parallax}}
\end{figure*}

We measured the parallax and proper motion of the masers relative to the two QSOs by differencing
their positions over time (see \citealt{Reid2009a} for a detailed discussion).  For the purposes of a maser
reference position, we used the spot at $-54.3$~\kms\ associated with feature No.~5 (in Table~\ref{tab3},
it is reported the peak LSR velocity for each feature). This spot was isolated,
nearly point-like, and fairly strong at all epochs ($\rm SNR \gg 100$).
Table~\ref{tab2} and Figure~\ref{parallax} show the results of the parallax and proper motion fitting for both
background sources.

Since systematic errors usually dominate over random noise, we estimated ``a posteriori'' the quality of the
measurements from the fit residuals (see \citealt{Reid2009a}). While both calibrators had a small angular
distance from the maser source (see Table~\ref{tab1}), J2018+3812 had a deconvolved size
(e.g., $\rm \approx 0.5~mas \times 0.2~mas$ at $-5\degr$, from the first epoch data) more than two times
that of J2015+3710. That affected the parallax signature in the north direction for the fainter calibrator
J2018+3812. Thus, to optimally combine the measurements of the two background sources, we
determined their relative weights as follow.  First, we fitted data for each background
source separately to give preliminary estimates of parallax and proper motion.
Since we expect no detectable proper motion for the extragalactic sources, we re-fitted the data
from the two sources independently, holding the proper motions fixed at the average values
from the first fit.  This procedure was iterated, adjusting the individual error floors to
yield values of chi-squared per degree of freedom near unity.
The error-floors determined in this manner were $\pm 0.01$~mas in both the E--W and N--S directions for J2015+3710
and $\pm 0.03$~mas in the E--W direction and $\pm 0.08$~mas in the N--S direction for J2018+3812.
These individual error floors were used in a final ``combined'' parallax reported in Table~\ref{tab2}.
Note that, since the targets of these observations were high-declination sources, we avoided low-elevation data.
That explains the accurate N--S offset measurements for the point-like calibrator J2015+3710, compared to
low-declination sources (e.g., \citealt{Reid2009a}). The parallax of \G\ from the combined fit is $0.108 \pm 0.005$~mas,
corresponding to a distance of $9.25^{+0.45}_{-0.40}$~kpc from the Sun.
For the IAU value of the distance to the Galactic center, R$_0 = 8.5$~kpc, our measured distance translates
to a Galactocentric radius of 10.8~kpc.

To determine a secular proper motion for the HMSFR (as opposed to a single maser spot), we need to correct the
combined fit values in Table~\ref{tab2} for the proper motion of the reference spot with respect to the forming star.
We estimated this contribution from the average proper motion of all the maser features with respect to the
reference spot of the parallax.
The average internal velocity components are $-0.28 \pm 0.09 $~mas~yr$^{-1}$ toward the east and
$-0.64 \pm 0.17 $~mas~yr$^{-1}$ toward the north, where we report the standard error of the mean.
Thus, the total motion of the whole source is estimated to be $-2.37 \pm 0.09 $~mas~yr$^{-1}$ and
$-4.48 \pm 0.17 $~mas~yr$^{-1}$ in the east and north directions, respectively.
At our measured distance these values correspond to $-104 \pm 4$~\kms\ and $-196 \pm 7$~\kms\ eastward and northward, respectively.
Completing the kinematic information, we have assumed an LSR velocity of $-57.0 \pm 2.2$~\kms\ for the HMSFR \G,
obtained from the CS(2$-$1) line survey by \citet{Bronfman1996}, in agreement with the median velocity of the maser
features ($-57.2$~\kms).

We can transform the 3-dimensional velocity from the equatorial heliocentric reference frame,
in which they were measured, to a reference frame rotating with the Galaxy, to estimate the peculiar motion
of the HMSFR (i.e. the deviation from a circular rotation).  Adopting a flat rotation curve for the Milky Way
with $\Theta = 239$~\kms\ \citep{Brunthaler2011}, a current ``best-estimate'' for the distance to the Galactic center
of R$_0 = 8.3$~kpc \citep{Reid2009b,Brunthaler2011}, and the revised \emph{Hipparcos} measurements of the solar
motion \citep{Schoenrich2010}, we find peculiar velocity components for \G\ of
$(U_s, V_s, W_s)=( 11.3 \pm 5.3, 1.1 \pm 8.8, -17.8 \pm 5.4)$~\kms, where U$_s$, V$_s$ and W$_s$ are directed toward
the Galactic center, in the direction of Galactic rotation and toward the North Galactic Pole, respectively.
In Table~\ref{outer}, we summarize these results together with the peculiar proper motions of sources previously
measured along the Outer arm.

\section{Discussion}

\begin{figure*}
\centering
\includegraphics[angle=0,scale= 0.9]{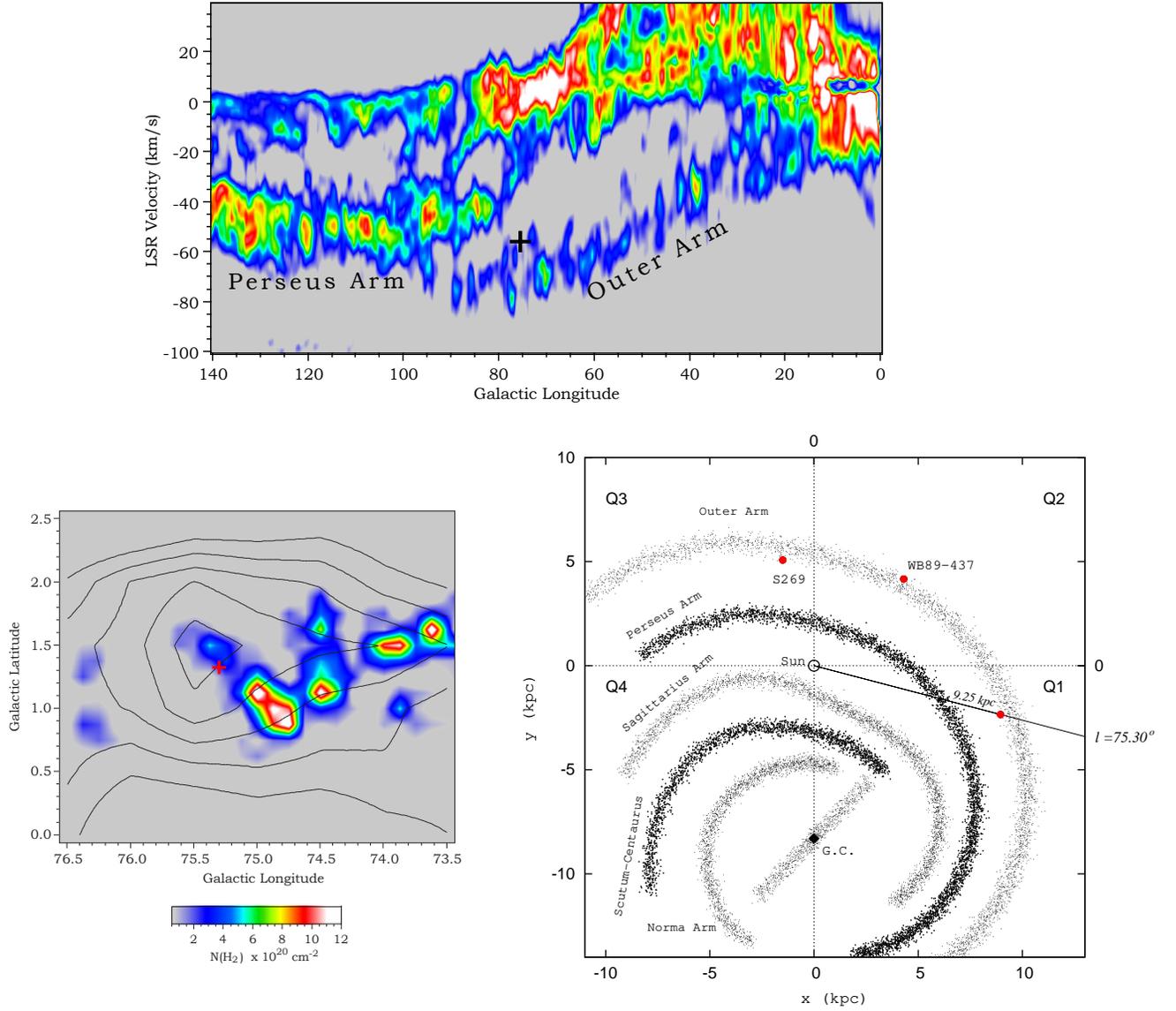}
\caption{{\scriptsize The star forming region G75.30+1.32 in the Galaxy, from the \HI\ and CO complex association to a
plane view of the Milky Way.
\textit{Upper Panel:} Longitude-velocity diagram of 21~cm emission at $b=0\degr$ from the LAB \HI\ survey
\citep{Kalberla2005}. The color scale from light blue to white corresponds to the 21~cm intensity range
35--120 K.  The emission from the Outer and Perseus Arms is labeled together with the position of G75.30+1.32 (cross).
\textit{Lower Left Panel:} Composite image of CO (colors) and \HI\ emission (contours) at the Galactic coordinates
of the source (cross). The CO emission from \citet{Dame2001} was integrated over the LSR velocity range from $-65$ to
$\rm -52~km~s^{-1}$. CO intensity was converted to $\rm H_2$ column density by using the relation
$\rm N(H_2)/W_{CO} = 1.8 \times 10^{20}$ \citep{Dame2001}. The \HI\ column density for the same velocity range starts
from $10^{20}$~cm$^{-2}$ in steps of $2 \times 10^{20}$~cm$^{-2}$.
\textit{Lower Right Panel:} Schematic view of the spiral arms of the Milky Way across the four Galactic quadrants after
\citet{Taylor1993} with updates. The best-value of R$_0 = 8.3$~kpc from \citet{Brunthaler2011} is assumed.
The location of the central bar from \citet{Benjamin2005} is also reported. The positions of the Outer Arm sources
previously measured with trigonometric parallaxes are labeled \citep{Reid2009b} together with the distance and Galactic
longitude of the present measurement.} \label{fig4}}
\end{figure*}

The source \G\ is located in the first Galactic quadrant at a Galactocentric radius of about 11~kpc, far beyond the
Galactocentric radius of the Perseus arm in this direction ($\sim 8$~kpc; lower right panel in Figure~\ref{fig4}).
The Outer arm between Galactic longitudes of $20\degr-90\degr$ at zero latitude can be clearly traced
in 21~cm \HI\ emission with $\rm V_{LSR} \leqslant -20$~\kms\ (upper panel of Figure~\ref{fig4}). The longitude-velocity
diagram in Figure~\ref{fig4}  shows that \G\ (marked by a cross) is definitely within
the locus of the Outer Arm, although it lies about 10 \kms\ higher in velocity than the central velocity of the arm at that
longitude ($\sim -67$~\kms). We note that, with a Galactic latitude of $b=1.32\degr$ and a distance of 9.25~kpc, \G\ is
located $\sim 213$~pc above the Galactic plane.  An analysis of both the CO survey from \citet{Dame2001} and the
\HI\ survey from \citet{Kalberla2005} toward the Galactic latitude of \G\ shows that the star-forming region sits on the
side of a fairly bright CO cloud and lies close to a substantial \HI\ enhancement (lower left panel in Figure~\ref{fig4}).
A further comparison of the Galactic coordinates and distance of \G\ with a vertical cross-section of the Galactic plane
points out the coincidence with a prominent warp of the Milky Way, particularly pronounced in the neutral hydrogen gas (e.g.,
\citealt{Nakanishi2003}).

The Galactic warp is a displacement of the outer disk from the midplane, generally upward (toward the North Galactic
Pole) between the first and second quadrants and downward at diametrically opposite directions in the third and
fourth. This midplane displacement can be also detected in faint CO line emission and peaks close to the Galactic
location of \G, at a Galactocentric azimuth $\theta=124\degr$, toward the North Galactic pole (e.g., \citealt{Nakanishi2006},
their Figure~8). The Galactocentric azimuth is defined as the azimuth angle around the rotation axis of the Galaxy,
taken so that $\theta=180\degr$ points to the Sun and $\theta=90\degr$ is parallel to $\ell=90\degr$. The difference of $\rm V_{LSR}$
at $b =0\degr$ and $b =1.32\degr$ may be related to the dynamics of the warp together with the peculiar motion of \G\
perpendicular to the plane ($-18$~\kms) and directed toward the Galactic plane. We note that this last value differs by
about six times the mean value for HMSFRs located within the Perseus arm ($\sim -3$~\kms; \citealt{Reid2009b}).

Two ideas have been forwarded to explain the Galactic warp (see \citealt{Sellwood2010} and reference therein):
1) as a consequence of the ongoing accretion of matter by the Galactic halo at much larger radii than the edge
of the disk and 2) as a result of tidal interactions due to the proximity of the Magellanic Clouds ($\ell \sim 280\degr$)
and the Sagittarius dwarf galaxy (e.g., Figure~1 in \citealt{Bailin2003}). In the accretion scenario, gravitational
torques between the inner halo and the outer disk produce a misalignment of the outer parts of the Galaxy
with respect to the plane of the inner disk. This happens because of a redistribution of angular momentum
through the system as long as freshly accreted matter streams from the outer to the inner parts of the Galaxy (e.g.,
\citealt{Binney1992,Jiang1999}). In this framework, the present result ($W_s$) may be interpreted as an inward motion
through the Galactic warp toward the inner Galactic disk.

Using R$_0 = 8.3$~kpc, $\Theta_0 = 239$~\kms, and a flat rotation curve beyond the solar circle, the \emph{revised} kinematic distance \citep{Reid2009b} at the Galactic coordinates of the source with an $\rm V_{LSR}=-57$~\kms\ is 9.9~kpc, with no kinematic distance
ambiguity as the source is in the outer Galaxy. The small discrepancy ($\sim1\sigma$) between the kinematic distance of \G\ and its
trigonometric distance of 9.25~kpc can be accounted for by the small peculiar motion of \G\ ($\sim7$~\kms) toward to the Sun direction
(Table~\ref{outer}). On the one hand, this small difference hints at the validity of a nearly flat rotation curve up to the outer
regions of the first Galactic quadrant ($d\Theta/dR < 3 \rm~ km~s^{-1}~kpc^{-1} $; e.g., \citealt{Brand1993,Reid2009b}).
On the other hand, the small values of the peculiar motion of \G\ along the Galactic plane (Table~\ref{outer}) may be interpreted
in the framework of a density-wave-induced spiral perturbation of the gravitational potential in a rotating galaxy \citep{Lin1969}.
These perturbations give rise to systematic, residual velocities within the spiral arms of the order of 10~\kms\
\citep{Lin1969}: inside the corotation radius, at the inner edge of an arm these streaming motions are directed
toward the galactic center and counter to galactic rotation; at the outer edge, residual velocities toward the galactic
center decrease to zero and point in the direction of the galactic rotation. Outside the corotation radius residual velocities
should be reversed (e.g., Figure~2 in \citealt{Mel'Nik1999}; \citealt{Mel'Nik2001}). By assuming a spiral pattern
speed of $\rm 20~km~s^{-1}~kpc^{-1} $ (e.g., \citealt{Bissantz2003}) and a Galactic rotation of about 240~\kms,
the corotation radius would be located at about 12~kpc. \G\ shows a residual velocity toward the Galactic center consistent with the
star-forming region inside the corotation radius (Table~\ref{outer}). Furthermore, the streaming motions of star-forming regions measured
so far along the Outer arm (Table~\ref{outer}) show positive values toward the Galactic center and suggest a corotation radius above a
Galactocentric distance of 13~kpc. On the contrary, \citet{Russeil2007} located the corotation radius between the Perseus arm and the
Outer arm in order to explain the opposite sign of residual velocities of their \HII\ regions. Note that, this estimate accounted
only for residual radial velocities ($\rm V_{LSR}$).  As soon as the BeSSeL survey\footnote{see the web page at the following URL,
http://www.mpifr-bonn.mpg.de/staff/abrunthaler/BeSSeL/index.shtml } and the VERA project\footnote{see the web page at the following URL,
http://veraserver.mtk.nao.ac.jp/index.html} increase the sample of 3-dimensional velocity measurements along the Outer arm,
we should be able to assess these issues with stronger statistical support.

\begin{figure*}
\centering
\includegraphics[angle=-90,scale=.6]{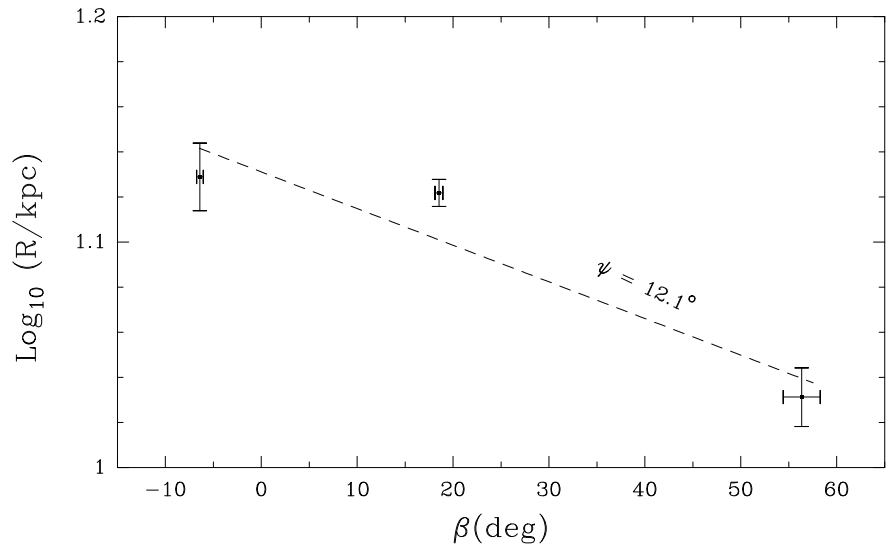}
\caption{Outer arm pitch angle ($\psi$). Fit of the Outer arm sources, located accurately by trigonometric parallaxes, with
an ideal model of logarithmic spiral arms (unweighted, dashed line). The logarithm of Galactocentric radius (R) is
plotted against Galactocentric longitude ($\beta$). The star-forming regions (dots) distribute about the mean position
of the Outer arm within an expected width of several hundreds of pc. \label{pitch}}
\end{figure*}

Finally, we can estimate the pitch angle of the Outer arm between Galactic longitudes from $75\degr$ to $196\degr$.
We mention as a caveat that spiral arms in other galaxies can have ``kinks'' and variations of pitch angle
over large portions of an arm (e.g., \citealt{Seigar1998}).
For the simplest case of a logarithmic spiral arm (e.g., \citealt{Reid2009b}), we fitted the parallax data of
\G\ (first quadrant) together with the two previous (trigonometric) measurements of star-forming regions along
the Outer arm (Table~\ref{outer}). This approach formally yields a pitch angle of $12.1\degr \pm 4.2 \degr$ which
covers about $120\degr$ of longitude between the first and third Galactic quadrants. In Figure~\ref{pitch}, we display
this result with the logarithm of the Galactocentric radius as a function of the Galactocentric
longitude ($\beta$, defined as 0 toward the Sun). While perfect-logarithmic, spiral arm sections should appear as straight lines
in such a plot, the measurement for the star-forming region WB~89--437 deviates from the fit by about three times
the parallax uncertainty ($\approx 600$~pc). This behavior may be explained when considering the spread of star-forming
regions within an arm with a width of several hundreds of pc.
The mean value derived here is in good agreement with a weighted average of published data for Galactic spiral
arms, which gives a mean pitch angle of $12.8\degr$ \citep{Vallee2008},
and agrees also with a more recent estimate of $13.5\degr$ for the Outer arm based on modeling of FIR cooling lines from the Galactic
interstellar medium  \citep{Steiman-Cameron2010}. Furthermore, it is consistent with the first direct estimate
of the Perseus arm pitch angle ($16.5\degr \pm 3.1 \degr$) made with the trigonometric distances of its HMSFRs \citep{Reid2009b}.

\acknowledgments
This work was partially funded by the ERC Advanced Investigator Grant GLOSTAR (247078).
Y.X. was supported by the Chinese NSF through grants NSF 11073054, NSF 10733030, and NSF 10621303.

{\it Facilities:} \facility{VLBA}.

\clearpage

\begin{deluxetable}{llllrcccc}
\tabletypesize{\scriptsize}
\tablecaption{Source Information\label{tab1}}
\tablewidth{0pt}
\tablehead{
\colhead{Source} & \colhead{R.A.~(J2000)} & \colhead{Decl.~(J2000)} & \colhead{$\theta_{\rm sep}$} & \colhead{P.A.} &
\colhead{HPBW} & \colhead{F$_{\rm peak}$} & Image rms &\colhead{V$_{\rm LSR}$} \\
\colhead{ }       & \colhead{(h m s)}       & \colhead{($\degr$ ' '')}    & \colhead{($\degr$)}     & \colhead{($\degr$)} &
\colhead{$ \rm mas \times mas \ at \ \degr$} & \colhead{(Jy beam$^{-1}$)} & (Jy beam$^{-1}$) &\colhead{(\kms)} \\}

\startdata

G75.30+1.32 & 20 16 16.0117 & 37 35 45.807 & \nodata & \nodata & $0.87\times0.43 \ at -15.9\degr$ & 4.499 & 0.007 &  $-56.8$  \\
J2015+3710  & 20 15 28.7297 & 37 10 59.514 & 0.4     & $-139$    & $0.88\times0.45 \ at -15.3\degr$ & 2.301 & 0.010 &  \\
J2018+3812  & 20 18 42.8500 & 38 12 41.700 & 0.8     & +58     & $0.89\times0.45 \ at -13.2\degr$ & 0.079 & 0.002 &         \\

\enddata
\tablecomments{Positions and source properties for the target maser and the QSO calibrators. The absolute position of the peak of the
phase-reference maser channel No.~131 (belonging to feature No.~2 in Table~\ref{tab3}) refers to the second (middle) epoch and is accurate
to within $\pm 1$~mas. We report the positions of the calibrators used in the VLBA correlator. Angular offsets ($\theta_{sep}$) and position
angles (P.A.) east of north relative to the maser source are indicated in columns 4 and 5. Columns 6, 7, and~8 give the natural restoring beam
sizes (HPBW), the peak intensities (F$_{\rm peak}$), and image rms noise of the phase-reference maser channel (at V$_{\rm LSR}$) and K-band
background sources from the second epoch.}

\end{deluxetable}

\begin{deluxetable}{lcccrrrr}
\tabletypesize{\scriptsize}
\tablecaption{Parameters of VLBA 22.2~GHz water maser features.\label{tab3}}
\tablewidth{0pt}
\tablehead{
\colhead{Feature} & \colhead{Detection} &\colhead{V$_{\rm LSR}$} & \colhead{F$_{\rm peak}$} &  \colhead{$\Delta \rm x$} & \colhead{$\Delta \rm y$} & \colhead{V$_{\rm x}$}    & \colhead{V$_{\rm y}$}  \\
\colhead{No.} & \colhead{(epochs)} &\colhead{(\kms)}    & \colhead{(Jy beam$^{-1}$)}    &  \colhead{(mas)} & \colhead{(mas)} &
\colhead{(mas yr$^{-1}$)} & \colhead{(mas yr$^{-1}$)} \\}

\startdata

1  &   \underline{2},3,4 & --53.89  &  9.84 &  $491.02 \pm 0.04 $ & $ -32.78 \pm 0.03 $  & $-0.11 \pm 0.11 $ & $ 0.54 \pm 0.04 $\\
2  & \underline{1},2,3,4 & --56.84  &  8.89 &  $     0 \pm 0.01 $ & $    0   \pm 0.01 $  &   $     0      $  &  $     0      $  \\
3  & \underline{1},2,3,4 & --56.42  &  7.87 &  $  1.37 \pm 0.03 $ & $  -0.91 \pm 0.03 $  & $-0.31 \pm 0.05 $ & $ 0.38 \pm 0.06 $\\
4  &   2,3,\underline{4} & --56.00  &  6.66 &  $  4.22 \pm 0.04 $ & $  -2.27 \pm 0.03 $  & $ 0.51 \pm 0.08 $ & $-0.41 \pm 0.04 $\\
5  & \underline{1},2,3,4 & --54.73  &  4.92 &  $297.69 \pm 0.03 $ & $ -32.02 \pm 0.01 $  & $ 0.59 \pm 0.03 $ & $ 0.70 \pm 0.01 $\\
6  & 1,2,3,\underline{4} & --55.57  &  4.80 &  $  3.51 \pm 0.02 $ & $  -1.95 \pm 0.01 $  & $ 0.05 \pm 0.09 $ & $ -0.07 \pm 0.05$\\
7  &   2,3,\underline{4} & --63.58  &  3.08 &  $304.16 \pm 0.05 $ & $ -38.64 \pm 0.03 $  & $ 0.62 \pm 0.06 $ & $ 0.28 \pm 0.03 $\\
8  &   2,3,\underline{4} & --58.52  &  2.54 &  $490.89 \pm 0.07 $ & $ -32.84 \pm 0.03 $  & $ 0.26 \pm 0.09 $ & $ 0.31 \pm 0.04 $\\
9  &   2,3,\underline{4} & --61.05  &  2.09 &  $530.13 \pm 0.09 $ & $  56.01 \pm 0.14 $  & $ 1.13 \pm 0.18 $ & $ 0.70 \pm 0.23 $\\
10 & \underline{1},2,3,4 & --55.57  &  1.50 &  $  2.71 \pm 0.02 $ & $  -1.50 \pm 0.02 $  & $ 0.00 \pm 0.04 $ & $ -0.38 \pm 0.02$\\
11 &   2,\underline{3},4 & --57.26  &  1.16 &  $540.14 \pm 0.04 $ & $  83.24 \pm 0.03 $  & $ 0.80 \pm 0.06 $ & $ 0.42 \pm 0.04 $\\
12 &     2,\underline{3} & --55.15  &  1.02 &  $532.29 \pm 0.04 $ & $  90.97 \pm 0.03 $  &        \nodata      &       \nodata  \\
13 & \underline{1},2,3,4 & --55.57  &  0.82 &  $298.30 \pm 0.06 $ & $ -32.21 \pm 0.03 $  & $ 0.86 \pm 0.06 $ & $ 0.63 \pm 0.06 $\\
14 & 1,2,\underline{3},4 & --56.00  &  0.70 &  $-74.65 \pm 0.02 $ & $ 224.93 \pm 0.04 $  & $ 0.57 \pm 0.04 $ & $ -1.39 \pm 0.07$\\
15 & 1,2,3,\underline{4} & --59.79  &  0.66 &  $240.94 \pm 0.04 $ & $-103.11 \pm 0.02 $  & $ 0.39 \pm 0.04 $ & $ 0.56 \pm 0.03 $\\
16 &   2,3,\underline{4} & --56.00  &  0.58 &  $  5.57 \pm 0.05 $ & $  -2.94 \pm 0.04 $  & $-0.03 \pm 0.09 $ & $ -0.43 \pm 0.07$\\
17 & 1                   & --48.41  &  0.55 &  $363.93 \pm 0.01 $ & $-110.72 \pm 0.01 $  &        \nodata      &       \nodata  \\
18 & 1,2,3,\underline{4} & --66.11  &  0.54 &  $304.20 \pm 0.02 $ & $ -38.95 \pm 0.04 $  & $ 0.47 \pm 0.02 $ & $ 0.37 \pm 0.03 $\\
19 & 4                   & --55.15  &  0.50 &  $-73.77 \pm 0.04 $ & $ 224.36 \pm 0.03 $  &        \nodata      &       \nodata  \\
20 & \underline{1},2,3,4 & --59.79  &  0.48 &  $-75.77 \pm 0.01 $ & $ 223.34 \pm 0.01 $  & $ 0.69 \pm 0.03 $ & $ -1.21 \pm 0.04$\\
21 & 4                   & --58.52  &  0.37 &  $535.22 \pm 0.02 $ & $  90.11 \pm 0.02 $  &        \nodata      &       \nodata  \\
22 &   \underline{2},3,4 & --52.63  &  0.32 &  $313.84 \pm 0.05 $ & $-635.91 \pm 0.04 $  & $ 0.15 \pm 0.12 $ & $ -0.38 \pm 0.09$\\
23 & 1                   & --54.73  &  0.32 &  $-74.32 \pm 0.01 $ & $ 226.09 \pm 0.03 $  &        \nodata      &       \nodata  \\
24 &   2,3,\underline{4} & --57.68  &  0.28 &  $-74.40 \pm 0.06 $ & $ 212.61 \pm 0.06 $  & $ 0.70 \pm 0.08 $ & $ -1.50 \pm 0.10$\\
25 & \underline{1},2,3,4 & --63.58  &  0.24 &  $302.19 \pm 0.01 $ & $ -33.75 \pm 0.02 $  & $ 0.18 \pm 0.03 $ & $ 0.17 \pm 0.05 $\\
26 & 4                   & --55.57  &  0.22 &  $489.33 \pm 0.03 $ & $ -34.53 \pm 0.03 $  &        \nodata      &       \nodata  \\
27 & 4                   & --62.32  &  0.22 &  $305.18 \pm 0.03 $ & $ -40.37 \pm 0.04 $  &        \nodata      &       \nodata  \\
28 &     \underline{2},3 & --54.31  &  0.20 &  $301.96 \pm 0.04 $ & $ -34.03 \pm 0.04 $  &        \nodata      &       \nodata  \\
29 &   2,3,\underline{4} & --58.10  &  0.18 &  $240.70 \pm 0.05 $ & $-102.68 \pm 0.04 $  & $-0.35 \pm 0.12 $ & $ 0.73 \pm 0.08 $\\
30 & 1                   & --50.52  &  0.16 &  $240.23 \pm 0.01 $ & $-103.19 \pm 0.02 $  &        \nodata      &       \nodata  \\
31 & 1                   & --54.73  &  0.16 &  $298.91 \pm 0.03 $ & $ -33.57 \pm 0.04 $  &        \nodata      &       \nodata  \\
32 & \underline{1},2,3   & --56.84  &  0.13 &  $  8.90 \pm 0.02 $ & $  -4.92 \pm 0.03 $  & $ 0.26 \pm 0.09 $ & $ -0.60 \pm 0.07$\\
33 & 3                   & --53.89  &  0.12 &  $  8.49 \pm 0.04 $ & $  -5.05 \pm 0.04 $  &        \nodata      &       \nodata  \\
34 & 1,2,3,\underline{4} & --59.79  &  0.11 &  $-83.93 \pm 0.03 $ & $ 242.82 \pm 0.05 $  & $-0.57 \pm 0.05 $ & $ 1.82 \pm 0.08 $\\
35 & 1                   & --58.94  &  0.10 &  $-76.87 \pm 0.01 $ & $ 231.80 \pm 0.02 $  &        \nodata      &       \nodata  \\
36 &     2,\underline{3} & --58.94  &  0.09 &  $-75.07 \pm 0.06 $ & $ 213.21 \pm 0.04 $  &        \nodata      &       \nodata  \\
37 &     2,\underline{3} & --61.47  &  0.07 &  $-73.28 \pm 0.05 $ & $ 211.22 \pm 0.05 $  &        \nodata      &       \nodata  \\
38 & 4                   & --50.52  &  0.07 &  $-83.85 \pm 0.06 $ & $ 246.97 \pm 0.07 $  &        \nodata      &       \nodata  \\

\enddata

\tablecomments{ For each identified feature, the label (given in Column~1) increases with decreasing brightness and the different epochs
of detection are reported in Column~2. Columns~3 and~4 report the LSR velocity and brightness of the brightest spot, observed at the epoch
underlined in Column~2. Columns~5 and~6 give the position offset relative to the feature No.~2 (centered on the phase-reference maser channel
No.~131, see Table~\ref{tab1}) in the east and north directions, respectively, at the first epoch of detection. The uncertainties give the
intensity-weighted standard deviation of the spots distribution within a feature. Columns~7 and~8 report the projected components of the
feature proper motion relative to the feature No.~2, along the east and north directions, respectively. A proper motion of 1~mas~yr$^{-1}$
corresponds to 44~km~s$^{-1}$ at a distance of 9.3~kpc.}

\end{deluxetable}

\begin{deluxetable}{lllll}
\tablecaption{G75.30+1.32:~Parallax \& Proper Motion Fits\label{tab2}}
\tablewidth{0pt}
\tablehead{
\colhead{Maser V$_{\rm LSR}$}   & \colhead{Background} & \colhead{Parallax} & \colhead{$\mu_{\rm x}$}         & \colhead{$\mu_{\rm y}$} \\
\colhead{(\kms)}     & \colhead{Source}     & \colhead{(mas)}    & \colhead{(mas yr$^{-1}$)} & \colhead{(mas yr$^{-1}$)} \\}

\startdata

$-54.3$ & J2015+3710 & $0.108 \pm 0.005$ & $-2.088 \pm 0.016 $ & $-3.847 \pm 0.019 $  \\
$-54.3$ & J2018+3812 & $0.096 \pm 0.008$ & $-2.153 \pm 0.023 $ & $-3.751 \pm 0.134 $  \\
    &         &                   &                     &                    \\
$-54.3$ & Combined & $ 0.108 \pm 0.005$ &  $-2.093 \pm 0.012 $ & $-3.845 \pm 0.014 $  \\
    &         &                   &                     &                    \\
\multicolumn{5}{c}{\textbf{Best-value for the Secular proper motion of G75.30+1.32}} \\
\hline\hline
    &         &                   &                     &                    \\
    &         & $ 0.108 \pm 0.005$&  $-2.37 \pm 0.09 $ & $-4.48 \pm 0.17 $  \\

\enddata
\tablecomments{Column 1 reports the LSR velocity of the reference maser spot (belonging to feature No.~5 in Table~\ref{tab3}); column 2
indicates the background sources whose data were used to model the relative proper motion of the maser for the parallax fit; column 3
reports the fitted parallax; columns 4 and 5 give the fitted proper motions along the east and north directions, respectively. Combined
fit used a single parallax parameter for the maser source position w.r.t. the two QSOs, after fixing the best overall proper motion in
each coordinate (see Section~\ref{distance_results}). The uncertainty for each parallax fit was obtained from the formal fitting
uncertainty. The Secular proper motion of \G\ was obtained by the combined fit value corrected for the internal proper motion of the
reference spot.}

\end{deluxetable}

\begin{deluxetable}{lrrcrrrc}
\tablecaption{Peculiar motion of sources along the Outer arm\label{outer}}
\tablewidth{0pt}
\tablehead{
\colhead{Source} & \colhead{$\ell$}  & \colhead{b}    & \colhead{R$_p$} & \colhead{U$_s$} & \colhead{V$_s$}    & \colhead{W$_s$} & Ref. \\
      & \colhead{(deg)} & \colhead{(deg)}& \colhead{(kpc)} & \colhead{(km s$^{-1}$)}& \colhead{(km s$^{-1}$)}& \colhead{(km s$^{-1}$)} & \\}

\startdata

G75.30+1.32  &  75.30 &  +1.32   & 10.7 & $ 11.3 \pm 5.3$ & $1.1 \pm 8.8 $  & $-17.8 \pm 5.4 $ &   \\
WB 89--437   & 135.28 &  +2.80   & 13.2 & $ 14.8 \pm 4.8$ & $2.4 \pm 8.3 $  & $0.9 \pm 9.8 $   & 1 \\
S 269        & 196.45 &  $-1.68$ & 13.4 & $ 5.0 \pm 3.5$  & $5.8 \pm 2.8 $  & $-4.4 \pm 1.5 $  & 2 \\
  &  &  &  &  &  & \\

\enddata
\tablecomments{Residual motion of sources that belong to the Outer arm with respect to a reference frame
rotating with the Galaxy and assuming a flat rotation curve. These fits consider the best estimates of R$_0 =8.3$~kpc and
$\Theta_0=239$~\kms\ given in \citet{Brunthaler2011} and the updated values of the solar motion in \citet{Schoenrich2010}.
Column 4 reports the projected Galactocentric distance of each source; column 5, 6, and 7 give the residual velocity components toward the
Galactic center, in the direction of Galactic rotation and toward the North Galactic Pole, respectively. Ref.: (1) \citet{Hachisuka2009};
(2) \citet{Honma2007}.
}

\end{deluxetable}

\end{document}